%% file: main.tex
\documentclass[conference]{IEEEtran}

\usepackage[utf8]{inputenc}
\usepackage[T1]{fontenc}
\usepackage{cite}
\usepackage{amsmath,amssymb,amsfonts}
\usepackage{algorithmic}
\usepackage{graphicx}
\usepackage{textcomp}
\usepackage{xcolor}
\usepackage{booktabs}
\usepackage{multirow}
\usepackage{bytefield}
\usepackage{hyperref}
\usepackage[scaled=1.0]{inconsolata}
\usepackage{microtype}
\usepackage{listings}
\usepackage{tikz}
\graphicspath{{./figures/}}

\begin{document}

\title{HyNoC: A Hybrid Circuit-Switch/Wormhole\\
Network-on-Chip for Distributed VLIW\\
Computing on FPGA}

\author{
  \IEEEauthorblockN{Christophe Clienti}
  \IEEEauthorblockA{%
    \textit{Independent Researcher}\\
    christophe.clienti@mines-paris.org\\
    \href{https://github.com/cclienti/verilog-ip/tree/master/hw/network/hynoc}{https://github.com/cclienti/verilog-ip/tree/master/hw/network/hynoc}}
}

\maketitle

\input{sections/abstract}
\input{sections/introduction}
\input{sections/related_work}
\input{sections/architecture}
\input{sections/switching}
\input{sections/target}
\input{sections/implementation}
\input{sections/evaluation}
\input{sections/conclusion}

\section*{Acknowledgment}
The author used Claude Sonnet~4.6 (Anthropic) as an AI writing assistant during
the preparation of this manuscript, including for drafting and revising text,
literature analysis, and \LaTeX{} editing. All technical content, claims, and
results have been verified by the author.

\bibliographystyle{IEEEtran}
\bibliography{references}

\end{document}

%% file: sections/abstract.tex
\begin{abstract}
Network-on-Chip (NoC) architectures have become the standard interconnect fabric
for many-core systems, yet most proposals face a fundamental trade-off between
latency, area, and congestion management. This paper presents \textbf{HyNoC}
(Hybrid Network-on-Chip), an open-source NoC architecture that combines
circuit-switch path establishment with wormhole data transfer, targeting
distributed computing systems built around VLIW processor cores on FPGA.

HyNoC employs source routing, where the complete path through the network is
encoded in the packet header by the sender or statically at compile time, enabling
both deterministic low-latency transfers and software-level hotspot avoidance
without the area overhead of virtual channels. The router features a parallel
round-robin arbiter (PRRA) with fixed grant latency, per-port independent clock
domains, and support for both unicast and multicast routing.

We discuss the design rationale, the hybrid switching model, and positioning
relative to prior NoC art, and argue that a richer topology combined with
compiler-assisted static routing is a competitive alternative to virtual channels
for FPGA-based distributed VLIW systems.  Verilator co-simulation measures the
deterministic per-hop latency and benchmarks the design at LLaMA~3~8B FFN
up-projection scale: a four-master quadrant configuration partitions the mesh
into traffic-isolated quadrants with zero cross-quadrant link traffic, yielding
a $5\times$ throughput improvement over a single master.  The complete RTL implementation is available as open-source hardware
under the CERN-OHL-P\,v2 license.
\end{abstract}

\begin{IEEEkeywords}
Network-on-Chip, FPGA, source routing, wormhole switching, circuit switching,
VLIW, parallel round-robin arbiter, open-source hardware.
\end{IEEEkeywords}

%% file: sections/introduction.tex
\section{Introduction}
\label{sec:introduction}

As the number of processing elements integrated on a single chip increases,
scalable on-chip communication becomes a first-class design concern.
Network-on-Chip (NoC) architectures have emerged as the standard answer to this
challenge, replacing shared buses and point-to-point links with a structured,
packet-switched interconnect fabric~\cite{dally2001}.

Existing NoC designs span a broad design space. Wormhole switching minimizes
buffering by forwarding flits as soon as the header is decoded, but exposes the
network to head-of-line blocking. Virtual channels (VCs) were introduced to
mitigate this blocking~\cite{dally1992}, at the cost of duplicated buffers that
can represent the dominant area contribution of a router~\cite{mello2005}. Circuit
switching provides fully dedicated, contention-free paths but requires path
reservation before any data can flow.

\textbf{HyNoC} takes a different approach: it combines the path-dedication
property of circuit switching---through a source-routed request/grant handshake at
each hop---with the flit-level, FIFO-backed flow control of wormhole switching.
This hybrid model is particularly well suited to FPGA targets, where block RAMs
used as FIFOs are scarce resources, and to distributed VLIW computing workloads,
where communication patterns are largely known at compile time and can therefore be
routed statically to avoid congestion.

Originally inspired by the Hermes NoC~\cite{moraes2004}, HyNoC departs
significantly from its ancestor by introducing distributed arbitration, relative
source routing with multicast support, per-port asynchronous clock domains, and a
parallel round-robin arbiter that grants requests in a one or two cycles.

The contributions of this paper are:
\begin{itemize}
  \item A precise characterization of HyNoC's hybrid switching model and its
        implications for latency and area;
  \item A discussion of static route management as an alternative to virtual
        channels, supported by a combinatorial analysis of shortest-path counts
        in mesh and higher-dimensional topologies;
  \item An analysis of the scope and limitations of the no-VC design choice;
  \item An open-source RTL implementation targeting both simulation and FPGA
        synthesis, complemented by the Veriparse toolkit~\cite{veriparse} for
        source-to-source elaboration of advanced RTL constructs and faster
        simulation;
  \item Verilator co-simulation results measuring the deterministic, linear
        per-hop latency and demonstrating link-disjoint quadrant isolation under
        a LLaMA~3~8B-scale distributed GeMV workload.
\end{itemize}

The remainder of the paper is organized as follows.
Section~\ref{sec:related} reviews related work.
Section~\ref{sec:architecture} describes the HyNoC architecture.
Section~\ref{sec:switching} details the hybrid switching model and the rationale
for avoiding virtual channels.
Section~\ref{sec:target} discusses the target application domain.
Section~\ref{sec:implementation} presents implementation details.
Section~\ref{sec:evaluation} reports simulation results, and
Section~\ref{sec:conclusion} concludes.

%% file: sections/related_work.tex
\section{Related Work}
\label{sec:related}

NoC architectures span a broad design space, comprehensively covered
in~\cite{dally2004}; this section reviews the dimensions most directly relevant
to HyNoC's design choices.

\subsection{Switching Strategies}

Three main switching strategies are found in the NoC literature.
\emph{Store-and-forward} switching requires a complete packet to be buffered at
each router before forwarding, providing clean error isolation but imposing large
buffer requirements and high latency~\cite{dally2001}.
\emph{Wormhole switching}~\cite{dally1987} pipelines flits across routers,
dramatically reducing buffer depth requirements; latency depends on path length
and contention.
\emph{Circuit switching}~\cite{kumar2002} reserves a dedicated path before
transmission, eliminating mid-network contention at the cost of a setup overhead
proportional to route length before the first data flit is delivered.

Hybrid approaches have been explored. {\AE}thereal~\cite{goossens2005} combines
guaranteed-throughput circuit-switched slots with best-effort wormhole traffic on
the same physical links using time-division multiplexing. HyNoC takes a different
hybrid point: it establishes a dedicated path through a hop-by-hop request/grant
handshake (the circuit phase), then streams payload flits in wormhole fashion
along that reserved path (the data phase). Unlike {\AE}thereal, HyNoC does not
multiplex traffic in time: once a path is open, it is held exclusively for the
duration of the packet, and no other packet can pre-empt it on the same egress
port. The setup latency — one request/grant round per hop — is paid once per
packet and is bounded by the route length, which is acceptable for the long-burst
transfers typical of VLIW inter-core communication.

\subsection{Routing Algorithms}

XY deterministic routing~\cite{dally2001} is simple and deadlock-free on 2D
meshes but provides only a single path between any pair of nodes. Adaptive
routing~\cite{duato1993} improves load balancing but requires more complex router
logic and typically needs virtual channels to avoid deadlock.

\emph{Source routing} encodes the complete route in the packet header at the
sender. This approach, studied for NoCs in~\cite{liwei2007} and~\cite{mubeen2010},
shifts routing complexity from the router to the sending node or to an offline
compilation step. The router itself becomes simpler: it only needs to decode the
next hop from the current flit, forward it, and update the index. Source routing
naturally supports any topology without per-router routing table storage, and
allows the sender to choose among multiple paths. This compile-time route
assignment property is particularly attractive for VLIW many-core systems, where
communication patterns are statically known and can be scheduled to avoid
contention (see Section~\ref{sec:target}).

Source routing has also been adopted in production VLIW many-core NoCs: the
Kalray MPPA2-256 uses a wormhole NoC with feed-forward source routing, where the
complete route is encoded in the header flit and each router forwards based solely
on a local offset extraction~\cite{dinechin2017}. This production deployment
validates the compile-time route assignment model that HyNoC targets.

HyNoC adopts source routing with a compact \emph{relative} hop encoding: since a
packet cannot exit from the same port it entered, a router with $N$ ports offers
only $N-1$ egress choices, encoded in $\lceil\log_2(N-1)\rceil$ bits per hop.
This reduces both header overhead and crossbar complexity simultaneously.

\subsection{Virtual Channels}

Virtual channels (VCs), introduced by Dally~\cite{dally1992}, multiplex several
logical channels over a single physical link, providing additional routing paths
and breaking deadlock cycles. Their main drawback is area: each VC requires a
dedicated buffer, and~\cite{mello2005} shows that router area scales nearly
linearly with VC count. On FPGAs, this overhead is particularly significant
because both the application logic and the communication fabric draw from a shared
pool of Block RAM (BRAM) and distributed LUT RAM resources: adding VCs directly
reduces the memory budget available to the application.

Several works have explored VC reduction or elimination.
Turn-model routing~\cite{glass1992} achieves deadlock freedom without VCs by
restricting the set of turns allowed in the routing graph. HyNoC's
approach---providing path diversity through richer topologies and compile-time
route selection---aligns with this trend, and is analyzed quantitatively in
Section~\ref{sec:switching}.

\subsection{NoC Arbitration}

Arbiter design is a central concern in input-queued routers, where multiple
ingress ports compete for the same egress~\cite{dally2004}.
Matrix arbiters evaluate all (ingress, egress) request pairs in parallel and can
grant in a single cycle regardless of port count, but require $O(N^2)$ state and
combinational logic, making them area-expensive for large routers.
Separable allocators decompose the two-dimensional matching problem into two
independent one-dimensional arbitration steps (one per ingress, one per egress),
reducing area at the cost of potentially requiring multiple rounds to reach a
stable matching~\cite{dally2004}.
Sequential round-robin FSMs scan requests cyclically, introducing up to $N-1$
cycles of latency before granting a pending request.

HyNoC implements a \emph{parallel round-robin arbiter} (PRRA): a set of $N-1$
combinatorial LUT-based state functions that evaluate all pending requests
simultaneously and jump directly to the winning state, regardless of how many
ports are idle. The PRRA matches the single-cycle grant latency of a matrix
arbiter for the common case, while its critical path does not grow with $N$ since
each LUT is evaluated independently. Fairness is maintained through a rotating
priority identical to classical round-robin.

\subsection{FPGA-Targeted NoCs}

FPGA implementations impose constraints not present in ASIC designs: Block RAMs
(BRAMs) are a finite, shared resource, so buffer-heavy NoC designs directly reduce
the memory budget available to the application; clock domain crossing between the
NoC fabric and the attached processing elements must be handled explicitly; and
routing resources constrain achievable clock frequencies~\cite{papamichael2012}.

CONNECT~\cite{papamichael2012} provides a parameterized, synthesizable NoC
generator for FPGAs supporting configurable topologies, VC counts, and buffer
depths, demonstrating that general-purpose FPGA NoCs are feasible. Its
VC-inclusive design trades BRAM for throughput robustness, which is appropriate
when traffic patterns are unpredictable at design time.

Hoplite~\cite{kapre2015} introduced deflection routing on FPGA overlays, achieving
very small area by eliminating input buffers entirely and routing contending flits
onto an alternative output rather than buffering them. HopliteRT~\cite{wasly2017}
extended Hoplite with a real-time traffic class, but deflection fundamentally
introduces non-deterministic, potentially unbounded delivery latency for best-effort
traffic — a property incompatible with the timing constraints of VLIW inter-core
communication.

Xilinx's Versal ACAP~\cite{swarbrick2019} embeds a hardened, fully programmable
NoC in silicon alongside the FPGA fabric, demonstrating that on-chip network
infrastructure has become a first-class concern even in commercial FPGA products.
Its hardened implementation achieves high bandwidth with low latency, but is
vendor-specific and not accessible as a soft IP.

Hermes~\cite{moraes2004}, the direct ancestor of HyNoC, demonstrated low-area
wormhole switching on FPGA with a 2D mesh topology and XY routing. Its
asynchronous variant Hermes-A~\cite{pontes2011} explored per-port independent
clock domains through asynchronous handshaking. HyNoC addresses the same
requirement using synchronous dual-clock FIFOs, which bridge arbitrary clock
domain boundaries at each ingress port without requiring asynchronous logic.

HyNoC targets a distinct point in this space: topology-agnostic source routing
with bounded latency, no virtual channels, and LUT-RAM-based FIFOs that avoid
BRAM consumption, suited to predictable VLIW workloads.

\subsection{VLIW Many-Core Systems}

Distributed VLIW many-core systems---exemplified by the Kalray
MPPA~\cite{dinechin2013}---replicate simple, compiler-scheduled
cores~\cite{ellis1986} connected through a dedicated NoC. Later
generations~\cite{dinechin2017} adopt source routing to preserve deterministic
latency guarantees across the fabric, the same design point HyNoC occupies.
Because the communication graph of such systems is known at compile time, the
bounded circuit-establishment latency of HyNoC's hybrid model (one PRRA grant
per hop) feeds directly into the compiler's timing model. We develop this
target application in detail in Section~\ref{sec:target}.

\subsection{Positioning}

Table~\ref{tab:positioning} summarizes how HyNoC relates to the principal design
axes discussed above.

\begin{table}[h]
\centering
\footnotesize
\caption{Positioning of representative NoCs}
\label{tab:positioning}
\resizebox{\columnwidth}{!}{%
\begin{tabular}{@{}lllll@{}}
\toprule
Design       & Switching  & Routing    & VCs & Target \\
\midrule
Hermes~\cite{moraes2004}      & Wormhole        & XY (dest.)    & No  & ASIC/FPGA \\
{\AE}thereal~\cite{goossens2005}  & TDM + Wormhole  & Table         & Yes & ASIC \\
CONNECT~\cite{papamichael2012}& Wormhole        & Configurable  & Yes & FPGA \\
HopliteRT~\cite{wasly2017}     & Deflection      & XY (dest.)    & No  & FPGA \\
HyNoC                         & Circuit+Wormhole& Source (rel.) & No  & FPGA \\
\bottomrule
\end{tabular}%
}
\end{table}

%% file: sections/architecture.tex
\section{HyNoC Architecture}
\label{sec:architecture}

\subsection{Overview}

A HyNoC network is built by assembling routers with a configurable number of ports
$N$ (3 to 9 in the current implementation), where each flit is $K+1$ bits wide
($K = \mathtt{PAYLOAD\_WIDTH}$, the extra bit being the stop bit). Each router port is full-duplex, composed
of an ingress interface and an egress interface. Two neighboring routers are
connected by cross-linking their egress and ingress interfaces. Processing nodes
are attached through a \emph{local interface}, which adds an extra output FIFO to
decouple the node's clock domain from the router.

\begin{figure}[t]
  \centering
  \includegraphics[width=\columnwidth]{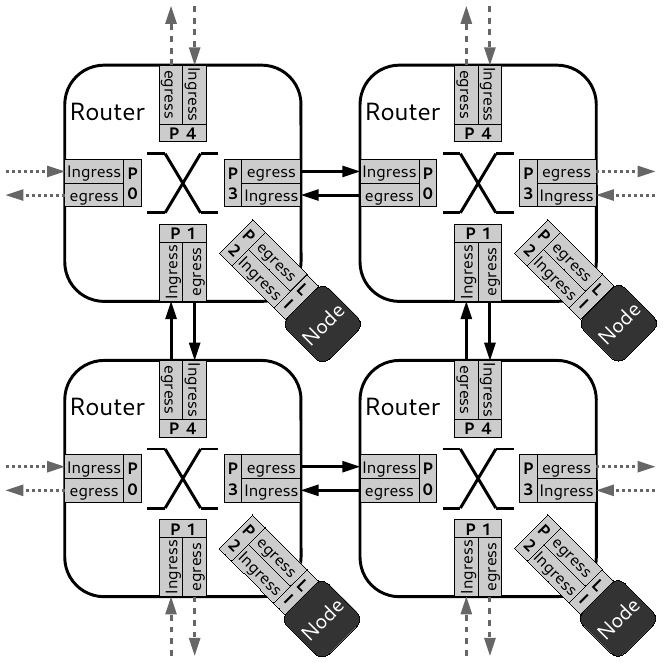}
  \caption{Router interconnections overview.}
  \label{fig:noc-overview}
\end{figure}

The router implements a \textbf{full crossbar}: any ingress port can simultaneously
communicate with any egress port, subject to arbitration. Multiple paths can be
open concurrently inside the same router, allowing concurrent transfers between different ingress-egress pairs.
Figure~\ref{fig:noc-overview} shows the internal organization of 5-port routers and their interconnections.

\subsection{Ingress Port}
The ingress port receives flits from the egress of a neighboring router. It buffers
incoming flits in a dual-clock FIFO ($2^D$ entries of $K+1$ bits), which
simultaneously handles clock domain crossing between the upstream router's clock
domain and the local router domain.

Once flits are buffered, the FSM inspects the two most significant bits of the
first flit to identify the flit type (routing or payload). It then extracts the
next-hop egress port identifier, asserts the matching \emph{request} bit among the
$N-1$ available egress ports, and waits for the corresponding \emph{grant}. Upon
grant, the ingress forwards the routing flit updated to the next hop (index
decremented); if the index has already reached zero, the routing flit is consumed
locally and not forwarded. Subsequent payload flits are streamed directly to the
granted egress until the stop bit of the last flit is detected, at which point the
channel is released.

\begin{figure}[t]
  \centering
  \includegraphics[width=\columnwidth]{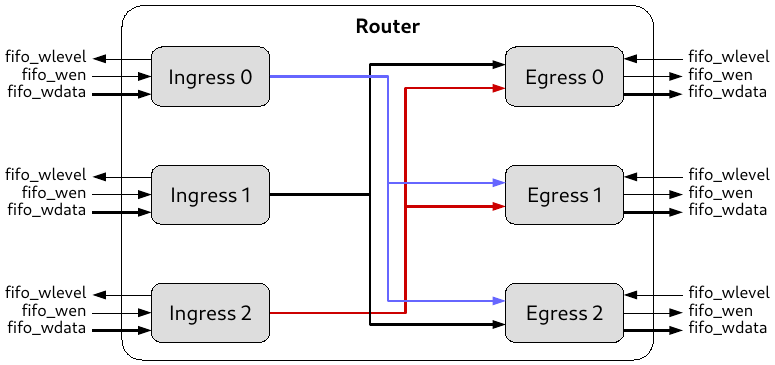}
  \caption{Internal data path of a 3-port router.}
  \label{fig:data-arch}
\end{figure}

Flow control is managed by monitoring both the downstream router's ingress FIFO
level and the local ingress FIFO level. The ingress stalls when either FIFO
approaches full, preventing flit loss.
The internal data path of a 3-port router is illustrated in Figure~\ref{fig:data-arch}.

\begin{figure}[t]
  \centering
  \includegraphics[width=\columnwidth]{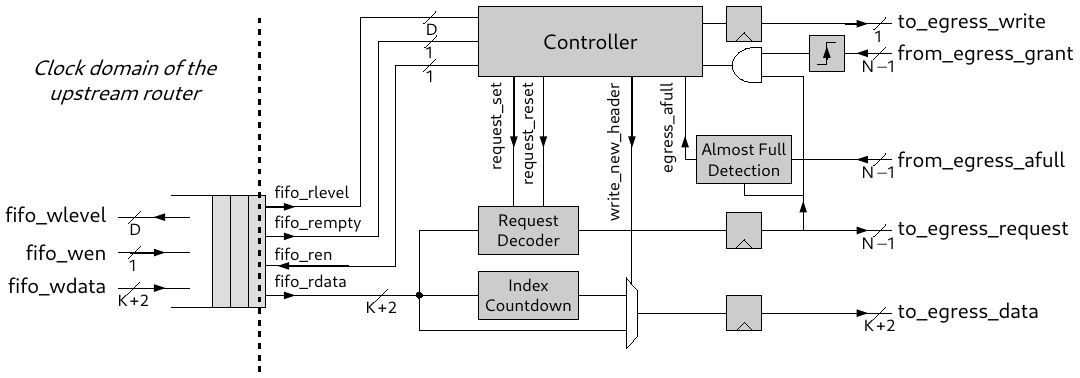}
  \caption{Ingress port internal architecture: dual-clock FIFO and routing FSM.}
  \label{fig:ingress-arch}
\end{figure}

\subsection{Egress Port}

\begin{figure}[t]
  \centering
  \includegraphics[width=\columnwidth]{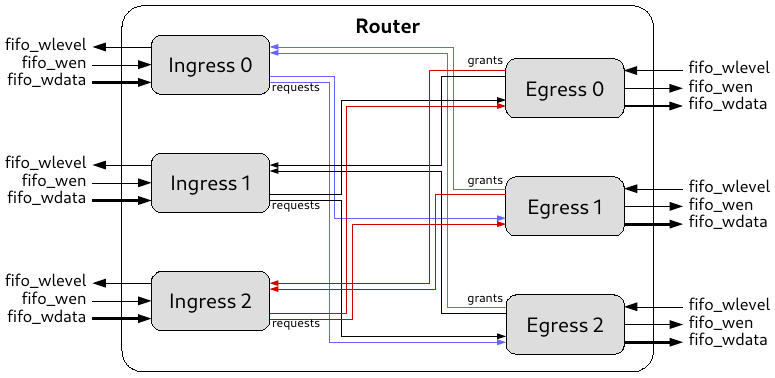}
  \caption{Internal control path of a 3-port router.}
  \label{fig:ctrl-arch}
\end{figure}

The egress port is connected to $N-1$ ingress ports via a $(K+1)$-bit wide data
path. It arbitrates among concurrent write requests using the PRRA, which responds
in one cycle when the port is idle, and grants one ingress at a time without
starvation. Once a grant is issued, the data and write-enable signals from the
winning ingress are routed directly to the downstream router's ingress FIFO. The
downstream FIFO level is forwarded back to the granted ingress for flow control.
The control path is shown in Figure~\ref{fig:ctrl-arch}: contrary to the data
path, control signals are point-to-point between each ingress and each egress,
avoiding any broadcast of control information.

\begin{figure}[t]
  \centering
  \includegraphics[width=\columnwidth]{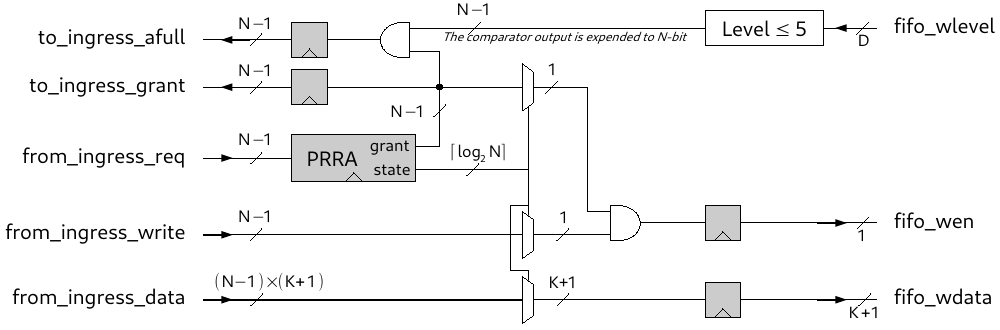}
  \caption{Egress port internal architecture: PRRA arbiter and data multiplexer.}
  \label{fig:egress-arch}
\end{figure}
\subsection{Parallel Round-Robin Arbiter (PRRA)}

\begin{figure}[t]
  \centering
  \includegraphics[width=\columnwidth]{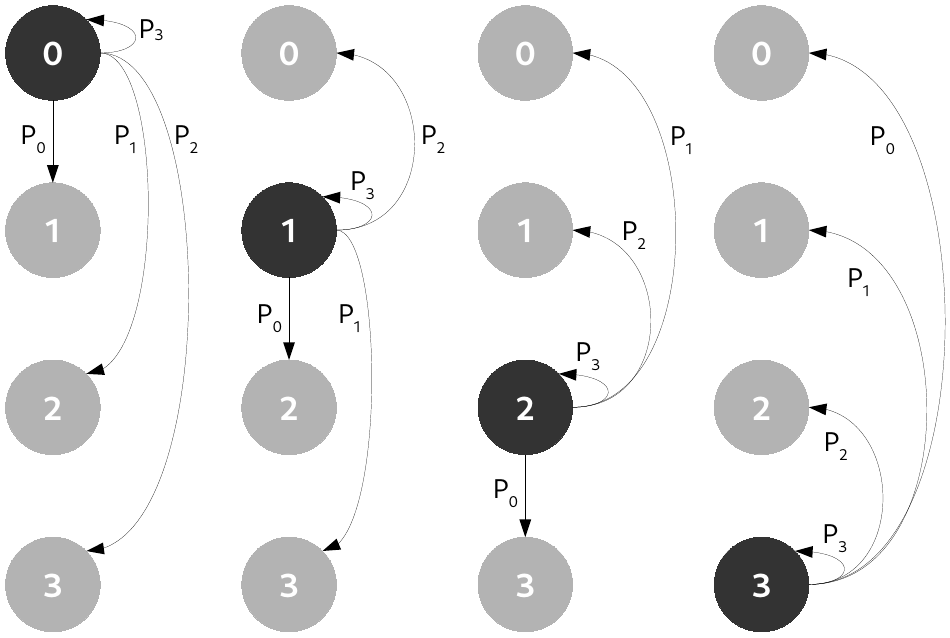}
  \caption{Parallel Round-Robin FSM: direct state transitions avoid scanning idle ports.}
  \label{fig:prra-fsm}
\end{figure}

The standard sequential round-robin FSM scans requests one by one in a fixed
order, introducing up to $N-1$ cycles of latency before granting a pending request
(Figure~\ref{fig:prra-fsm}). HyNoC replaces this with a \textbf{parallel
  round-robin arbiter (PRRA)}: a set of $N-1$ combinatorial LUT-based state functions
that evaluate all pending requests simultaneously and jump directly to the winning
state, regardless of how many ports are idle.

\begin{figure}[t]
  \centering
  \includegraphics[width=\columnwidth]{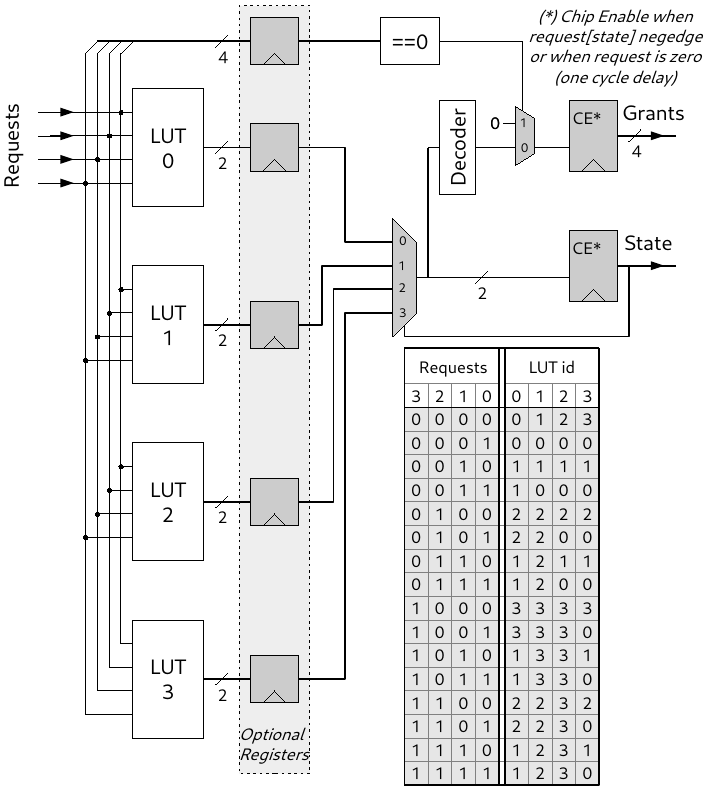}
  \caption{PRRA architecture: $N-1$ parallel LUTs, one per priority offset.}
  \label{fig:prra-arch}
\end{figure}

The architecture instantiates $N-1$ \texttt{prra\_lut} modules in parallel
(Figure~\ref{fig:prra-arch}), each pre-computing the next granted port for a
different starting priority offset. The LUT corresponding to the current state is
selected by a multiplexer; its output feeds the state register. The state
transitions only when the currently granted request is de-asserted (end of packet)
or when no request is active, locking the grant for the entire packet transfer and
preventing spurious re-arbitration.

The arbiter responds in \textbf{one or two clock cycles} depending on the
\texttt{PIPELINE} parameter: without pipelining the LUTs are purely combinatorial
and the grant is issued in one cycle; with pipelining an extra register stage
on the LUT outputs reduces the critical path at the cost of one additional cycle.
The critical path does not grow with $N$ because each LUT is independent and
selected, not chained. Starvation is prevented by the rotating priority: after
serving port $k$, priority rotates to port $k+1 \mod N$.

\subsection{Clock Domain Architecture}

Each router port operates in its own clock domain. The dual-clock FIFO at each
ingress interface handles the crossing between the upstream port clock and the
local router clock. For timing-critical designs, a single-clock mode
(\texttt{SINGLE\_CLOCK\_ROUTER=1}) is available, reducing crossing latency.

\subsection{Local Interface}

Any router port can be connected to a processing node instead of a neighboring
router. The \texttt{hynoc\_local\_interface} module handles this attachment by
instantiating an additional FIFO at the egress output. This FIFO serves two
purposes: it fully decouples the node's clock domain from the router clock domain,
and it smooths traffic bursts, preventing the node from becoming a bottleneck when
it cannot consume data fast enough. From the router's perspective, the local
interface is indistinguishable from a downstream router's ingress port, preserving
the uniformity of the flow control mechanism across all port types.

\textbf{AXI-Stream compatibility.} The port-level handshake is deliberately
modeled after the AXI-Stream protocol~\cite{arm_axi}: the \texttt{write} signal
corresponds to \texttt{TVALID}, the \texttt{full} signal to the negation of
\texttt{TREADY}, the data bus to \texttt{TDATA}, and the stop bit to
\texttt{TLAST}. This mapping makes the interface immediately familiar to hardware
designers working in the FPGA ecosystem. The only extension beyond standard
AXI-Stream is the \texttt{fifo\_level} sideband, which exposes the downstream
buffer occupancy to allow the sender to anticipate back-pressure before the
\texttt{full} condition is reached.

\subsection{First Layer Protocol}

\subsubsection{Packet and Flit Structure}

\begin{figure}[t]
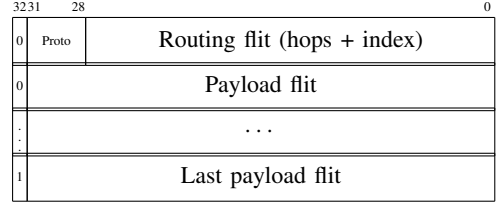

  \centering
  \begin{bytefield}[bitwidth=0.55em, endianness=big]{33}
    \bitheader{32,31,28,0} \\
    \bitbox{1}{\tiny 0} & \bitbox{4}{\tiny Proto} & \bitbox{28}{\small Routing flit (hops + index)} \\
    \bitbox{1}{\tiny 0} & \bitbox{32}{\small Payload flit} \\
    \bitbox{1}{\tiny $\vdots$} & \bitbox{32}{\small $\cdots$} \\
    \bitbox{1}{\tiny 1} & \bitbox{32}{\small Last payload flit} \\
  \end{bytefield}
  \caption{Packet structure: routing flit(s) followed by payload flits.
    The MSB is the stop bit (0=intermediate, 1=last flit).}
  \label{fig:pkt-struct}
\end{figure}

\begin{figure}[t]
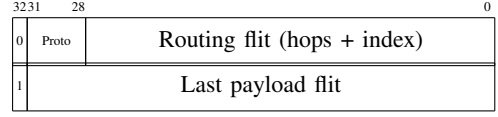

  \centering
  \begin{bytefield}[bitwidth=0.55em, endianness=big]{33}
    \bitheader{32,31,28,0} \\
    \bitbox{1}{\tiny 0} & \bitbox{4}{\tiny Proto} & \bitbox{28}{\small Routing flit (hops + index)} \\
    \bitbox{1}{\tiny 1} & \bitbox{32}{\small Last payload flit} \\
  \end{bytefield}
  \caption{Minimal HyNoC packet: one routing flit (stop bit = 0) followed by
    one payload flit (stop bit = 1). The routing flit stop bit must be
    zero since the ingress FSM rejects routing flits with stop bit set.}
  \label{fig:pkt-min}
\end{figure}

The network operates on flits of $K+1$ bits ($K = \mathtt{PAYLOAD\_WIDTH}$). The
most significant bit is a \emph{stop bit}: when set, it marks the last flit of a
packet. A packet is built with at least one \emph{Network Hops flit} (routing
flit) followed by at least one \emph{payload flit}, as shown in
Figure~\ref{fig:pkt-struct}. If multiple routing flits are present, intermediate
routers treat all but the first as payload; each router consumes one routing flit
from the front of the chain. The MSB of a routing flit is always zero, reserving
space for future sub-type extensions. Figure~\ref{fig:pkt-min} shows the smallest
valid packet: one routing flit (stop bit must be zero) followed by one payload flit.

The routing flit carries a 4-bit \emph{Proto} field (Table~\ref{tab:proto}):

\begin{table}[h]
  \centering
  \footnotesize
  \caption{Supported routing protocols}
  \label{tab:proto}
  \begin{tabular}{@{}ll@{}}
    \toprule
    Proto & Routing method \\
    \midrule
    \texttt{4'b0000} & Unicast circuit switch \\
    \texttt{4'b0001} & Multicast circuit switch \\
    \texttt{4'b1000} & XY routing (slot reserved, not yet implemented) \\
    \texttt{4'b1111} & Forbidden \\
    \bottomrule
  \end{tabular}
\end{table}

The proto value \texttt{4'b1000} is reserved for future XY routing support.
XY routing would allow destination-addressed packets using (x,y) coordinates
rather than explicit hop lists, enabling simpler packet construction at the
cost of restricting the topology to 2D meshes. The ingress FSM currently
flushes any packet carrying this proto value.

\subsubsection{Unicast Source Routing}

\begin{figure}[t]
  \centering
  \includegraphics[width=0.85\columnwidth]{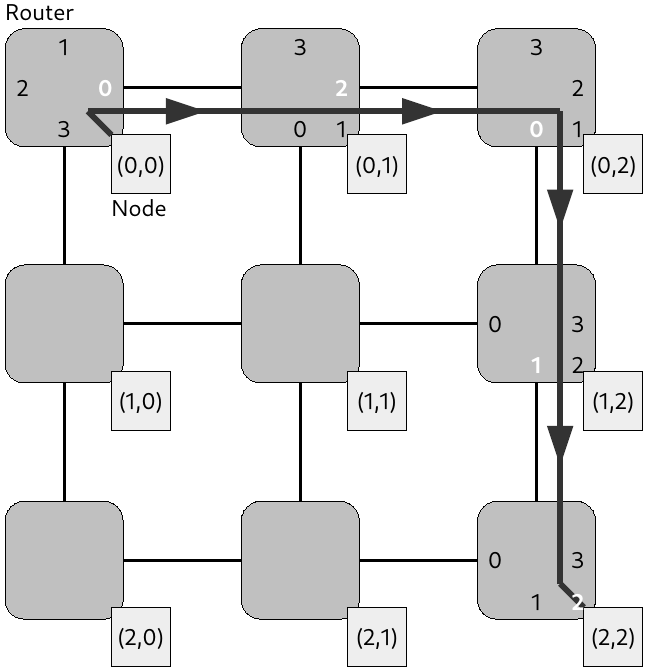}
  \caption{Relative hop encoding for a $3\times 3$ mesh network built with 5-port
    routers. The hop list $[0 \to 2 \to 0 \to 1 \to 2]$ establishes a
    path from node $(0,0)$ to node $(2,2)$. Each hop ID is relative to
    the ingress port: hop~0 denotes the next counterclockwise egress,
    excluding the ingress port itself.}
  \label{fig:hops}
\end{figure}

The Network Hops flit structure is shown in Figure~\ref{fig:net-hops}. It carries
$H$ hop fields of $\lceil\log_2(N-1)\rceil$ bits each, encoding the relative
egress port ID at the next router, together with an \emph{index} field initialized
to $H-1$. The index directly feeds the multiplexer selecting the active hop in the
ingress port. It is decremented after each hop is consumed; if the index is already
zero before path opening, the routing flit is not forwarded downstream. The relative hop encoding on a $3\times3$ mesh of 5-port routers is
illustrated in Figure~\ref{fig:hops}.

\begin{figure}[t]
  \centering
  \begin{bytefield}[bitwidth=0.75em, endianness=big]{33}
    \bitheader{32,31,28,27,26,5,4,3,0} \\
    \bitbox{1}{\tiny 0} &
    \bitbox{4}{\small Proto} &
    \bitbox{2}{\small H11} &
    \bitbox{20}{\small Hops 10 -- 1} &
    \bitbox{2}{\small H0} &
    \bitbox{4}{\small Index} \\
  \end{bytefield}
  \caption{Network Hops flit structure for a 5-port, 32-bit router (unicast,
    12 hops of 2~bits, 4-bit index). The layout from MSB to LSB is:
    stop bit, Proto, \emph{Gap} (if any), Hop $H-1$ \ldots Hop~0, Index.
    The Gap absorbs unused bits between the Proto field and the first hop,
    and is zero in this particular configuration.
    The \emph{Index} field (initialized to $H-1$) selects the active hop.}
  \label{fig:net-hops}
\end{figure}

Since a packet cannot exit through the same port it entered, each ingress port has
only $N-1$ reachable egress ports. Hop fields therefore require only
$\lceil\log_2(N-1)\rceil$ bits, which yields particularly compact encodings for
the most common router sizes: a 3-port router encodes hops on 1~bit, a 5-port
router on 2~bits, and a 9-port router on 3~bits. This reduction simultaneously
shrinks the routing flit header overhead and the internal crossbar size, since the
crossbar itself does not need to include the loopback path from ingress to its
paired egress.

More precisely, the encoding constrains $N$ to satisfy $2^{\lceil\log_2(N-1)\rceil} + 1 = N$,
meaning only $N \in \{3, 5, 9\}$ are valid port counts. This constraint is enforced
at elaboration time in the RTL. The relative hop ID is computed counterclockwise
from the ingress port: hop~0 always refers to the next counterclockwise egress,
and the ingress FSM resolves the relative ID into a one-hot physical egress request.

Given a flit of $K+1$ bits, a 4-bit Proto field, and an \texttt{INDEX\_WIDTH}-bit
index field, the number of available hop bits is $K - 3 - \mathtt{INDEX\_WIDTH}$,
and the maximum hop count is:
\begin{equation}
  H_\text{max} = \left\lfloor \frac{K - 3 - \mathtt{INDEX\_WIDTH}}
  {\lceil\log_2(N-1)\rceil} \right\rfloor
  \label{eq:hmax}
\end{equation}

Table~\ref{tab:hops} gives the resulting values for the three valid router sizes
in the default configuration ($K=32$, $\mathtt{INDEX\_WIDTH}=4$), for both
unicast and multicast routing (which uses $N-1$ bits per hop instead of
$\lceil\log_2(N-1)\rceil$). The wider multicast hop field halves the maximum
hop count relative to unicast, reflecting the inherent trade-off between
fanout expressiveness and path length.

\begin{table}[h]
  \centering
  \footnotesize
  \caption{Maximum hop count per routing mode ($K=32$, \texttt{INDEX\_WIDTH}=4)}
  \label{tab:hops}
  \begin{tabular}{@{}lrrrr@{}}
    \toprule
    Router & \multicolumn{2}{c}{Unicast} & \multicolumn{2}{c}{Multicast} \\
    & bits/hop & $H_\text{max}$ & bits/hop & $H_\text{max}$ \\
    \midrule
    3-port & 1 & 24 & 2 & 12 \\
    5-port & 2 & 12 & 4 &  6 \\
    9-port & 3 &  8 & 8 &  3 \\
    \bottomrule
  \end{tabular}
\end{table}
\subsubsection{Multicast Source Routing}

In multicast mode, each hop field is a bitmask of width $N-1$, where each bit
selects one egress port. The packet is duplicated and forwarded to all asserted
egress ports simultaneously. Multiple routing flits with different multicast masks
can be chained to address complex multicast trees. This is particularly useful for
broadcast operations in VLIW many-core workloads.

%% file: sections/switching.tex
\section{Hybrid Switching Model and Design Rationale}
\label{sec:switching}

\subsection{The Hybrid Switching Model}

HyNoC's switching model is best described as \emph{circuit-switched wormhole}: it
combines a \textbf{circuit-establishment phase} with a \textbf{wormhole data
phase}.

In the \textbf{circuit-establishment phase}, the ingress port asserts a request to
a specific egress port (identified from the routing flit) and waits for a grant.
This request/grant handshake propagates along the route: once the first router
grants the request, the ingress begins forwarding flits, which in turn trigger the
same handshake at the next router. A dedicated forwarding path through the network
is thus progressively opened, hop by hop.

In the \textbf{wormhole data phase}, payload flits are streamed flit-by-flit along
the established path, regulated by FIFO-level flow control. The path remains
dedicated to this packet until the stop bit is observed, ensuring in-order,
contention-free delivery once the channel is open.

This combination provides the \emph{low buffering} advantage of wormhole switching
(only the ingress FIFO is needed, no full packet storage) while offering the
\emph{path dedication} advantage of circuit switching (no mid-network arbitration
contention during payload transfer). The trade-off is a path-establishment latency
proportional to the route length before the first payload flit is delivered at the
destination.

\subsection{End-to-End Latency}

Let $H$ denote the number of hops from source to destination (including the
final local hop), and $P$ the number of payload flits in the packet.
End-to-end latency decomposes into two additive terms.

\emph{Setup phase.} The routing flit propagates hop by hop. At each hop it
traverses a fixed pipeline of register stages---the ingress FIFO read, the
routing-FSM request register, the PRRA grant, and the egress data
register---and, in dual-clock mode, the gray-code synchroniser of the ingress
FIFO~\cite{cummings2002}. This pipeline is identical for every hop and contains
no contention-dependent path once a request is granted, so the per-hop cost is
a fixed constant~$\alpha$.

\emph{Transfer phase.} Once the path is established, the $P$ payload flits are
pipelined at one flit per hop: the first flit reaches the destination after
$H$ cycles and the last after $H + P - 1$ cycles.

End-to-end latency is therefore strictly linear and deterministic:
\begin{equation}
  L = \alpha\,H + P - 1
  \label{eq:latency}
\end{equation}
with no congestion-dependent term once the path is open. The per-hop constant
$\alpha$ depends only on the FIFO clocking mode---it is larger in dual-clock
mode by the depth of the gray-code synchroniser---and is measured directly by
co-simulation in Section~\ref{sec:evaluation}.

\subsection{Static Route Management as an Alternative to Virtual Channels}

Virtual channels were introduced primarily to: (1) provide additional logical paths
to avoid deadlock; and (2) reduce head-of-line blocking under congestion. HyNoC
addresses both goals differently.

\textbf{Deadlock avoidance.} Deadlock in wormhole networks arises from cyclic
dependencies in the channel dependency graph (CDG)~\cite{dally1987}: a set of
flits each waiting for a channel held by the next forms an unbreakable cycle.
With source routing, the complete path is fixed at injection time by the sender.
Deadlock freedom therefore reduces to a property of the \emph{route set}: if
every injected route follows a routing function whose CDG is acyclic ---
dimension-order routing being the canonical example --- no cycle can form,
regardless of traffic load. This analysis is performed once, statically, by the
compiler or network configurator; the router itself carries no deadlock-prevention
logic. The guarantee is strictly stronger than VC-based approaches, where
deadlock freedom depends on runtime routing decisions and requires the router to
maintain channel state across multiple virtual queues.

\textbf{Congestion avoidance} is addressed at two levels. First, the compiler or
runtime system assigns routes across multiple available shortest paths, distributing
traffic away from known hotspots. The number of shortest paths between two nodes in
an $R \times C$ mesh is a classical combinatorial result~\cite{knuth1997}: each path consists of a fixed sequence of moves along each dimension, taken in any order, and the count is given by the multinomial coefficient:

\begin{equation}
  p = \binom{(R-1)+(C-1)}{R-1} = \frac{[(R-1)+(C-1)]!}{(R-1)!\,(C-1)!}
  \label{eq:paths2d}
\end{equation}

This generalizes to a $K$-dimensional network $n_1 \times n_2 \times \cdots \times
n_K$ as:

\begin{equation}
  p = \frac{\left[\sum_{i=1}^{K}(n_i-1)\right]!}{\prod_{i=1}^{K}(n_i-1)!}
  \label{eq:pathsnd}
\end{equation}

Note that~(\ref{eq:paths2d}) and~(\ref{eq:pathsnd}) give the number of shortest
paths between the \emph{most distant} node pair (corner-to-corner). For an
arbitrary pair separated by $(d_x, d_y)$ hops the count is $\binom{d_x+d_y}{d_x}$,
which is lower. Table~\ref{tab:paths} reports corner-to-corner values for
representative 2D mesh topologies.

\begin{table}[h]
\centering
\footnotesize
\caption{Shortest path counts in 2D mesh NoCs (corner-to-corner)}
\label{tab:paths}
\begin{tabular}{@{}lrrl@{}}
\toprule
Topology & Routers & Shortest paths & Path length \\
\midrule
$2\times 2$   &   4 &               2 &  2 \\
$4\times 4$   &  16 &              20 &  6 \\
$8\times 8$   &  64 &           3\,432 & 14 \\
$16\times 16$ & 256 & 155\,117\,520    & 30 \\
\bottomrule
\end{tabular}
\end{table}

Second, for applications with more dynamic traffic, the network topology can be
enriched with additional relay routers (without local interfaces), increasing path
diversity proportionally to the added area.

\subsection{Area Comparison with Virtual Channels}

Following~\cite{mello2005}, router area scales linearly with the number of VCs
(each VC requires an additional input buffer per port). An $8\times 8$ mesh with
4~VCs consumes the equivalent area of a $16\times 16$ mesh without VCs, yet the
latter offers over 45,000$\times$ more shortest paths between corner nodes
(Table~\ref{tab:paths}: 155,117,520 vs.\ 3,432). On FPGA, where
block RAMs are scarce, minimizing buffer count is a primary concern.

The VC area model is itself optimistic in two respects: (1) the crossbar logic and
VC scheduler also grow with VC count; (2) VC path diversity is an upper bound,
since in practice the VC is assigned at the source and remains fixed for the entire
path, providing less diversity than a true additional physical dimension.
Conversely, the physical router area model does not capture link wiring cost, which
may be relevant in ASIC contexts.

\subsection{Head-of-Line Blocking and Mitigations}

Once a channel is granted, it remains dedicated until the stop bit of the last
flit is received. A long packet therefore delays other packets competing for the
same egress port. Two complementary mitigations are available in HyNoC:

\begin{enumerate}
  \item \textbf{Network enrichment}: adding more routers distributes traffic
    across more egress ports, reducing the probability that multiple flows compete
    for the same port. The cost is increased end-to-end \emph{latency} due to
    additional hops---this is the primary performance trade-off, not throughput.
  \item \textbf{Multiple local interfaces}: a node can expose several local
    interfaces connected to different router ports, providing as many independent
    logical channels as needed. This replicates the path diversity benefit of VCs
    without duplicating internal router buffers. It is the recommended approach
    when a node requires simultaneous, non-blocking communication with multiple
    peers.
\end{enumerate}

\subsection{Scope and Limitations}

The topology-based approach is most effective when traffic patterns are known at
compile time and routes can be assigned statically. When traffic is entirely
unpredictable at design time and topology enrichment is not feasible, virtual
channels remain the more appropriate solution. HyNoC targets systems---distributed
VLIW processors on FPGA---where compile-time route assignment is natural and FPGA
resource budgets make buffer minimization a primary concern.

%% file: sections/target.tex
\section{Target Application: Distributed VLIW Computing}
\label{sec:target}

\subsection{VLIW Cores and Compile-Time Scheduling}

VLIW processors expose multiple functional units to the compiler, which schedules
operations explicitly in long instruction words~\cite{ellis1986}. This design
philosophy shifts control complexity from hardware to software, resulting in
simple, area-efficient cores well suited for replication in many-core arrays.
Signal processing, scientific computing, and neural network inference kernels are
representative workloads.

In a distributed VLIW many-core system, each core executes a portion of a
data-flow graph, passing intermediate results to neighboring cores through the
NoC. The communication graph is typically known at compile time, enabling the
compiler to:
\begin{enumerate}
  \item Assign each communication a source route through the NoC;
  \item Schedule packet injections to avoid simultaneous contention on the same
        egress port;
  \item Guarantee bounded communication latency for critical paths.
\end{enumerate}

This compile-time visibility makes HyNoC's hybrid model particularly effective:
the circuit-establishment latency is paid once per packet and is deterministic
(bounded by route length and PRRA grant latency), and the wormhole payload
transfer proceeds without further arbitration overhead.

\subsection{Integration with the Local Interface}

Each VLIW core connects to the NoC through a local interface, which presents a
simple FIFO-based write/read API. The core writes routing flits followed by
payload flits into the local interface's ingress FIFO; the local interface manages
all handshaking with the router. On the receive side, incoming flits are deposited
in the egress FIFO and read by the core at its own pace. The dual-FIFO structure
fully decouples the core's clock domain from the router domain.

The \texttt{hynoc\_stream\_writer} and \texttt{hynoc\_stream\_reader} modules are
simulation utilities that generate and check random packet streams, respectively.
They serve as traffic generators in the testbench environment, verifying the
correctness of the NoC under various load conditions.

\subsection{Multicast for Broadcast Operations}

VLIW many-core applications frequently require broadcast operations: distributing
a coefficient matrix, a lookup table, or a control token to multiple cores
simultaneously. HyNoC's multicast routing protocol supports this natively: a
single packet with a multicast routing flit is replicated at each router toward
all targeted egress ports, without requiring the sender to issue multiple unicast
packets. This reduces network load and sender overhead for one-to-many
communication patterns common in data-parallel VLIW workloads.

%% file: sections/implementation.tex
\section{Implementation}
\label{sec:implementation}

\subsection{Repository Organization}

HyNoC is part of a larger open-source IP library hosted at
\url{https://github.com/cclienti/verilog-ip}. The repository is organized into
two complementary layers under the \texttt{hw/} directory:

\begin{itemize}
  \item \texttt{hw/lib/}: a collection of reusable, standalone Verilog IP cores
        covering arithmetic, memories, FIFOs, and arbiters. Each module ships
        with its own parameterized RTL and self-checking testbench.
  \item \texttt{hw/network/hynoc/}: the HyNoC subsystem, which builds on top of
        \texttt{hw/lib/} and is described in detail in this paper.
\end{itemize}

HyNoC directly instantiates three library modules:
\texttt{dclkfifolut} (dual-clock FIFO mapped to distributed LUT RAM, used in
\texttt{hynoc\_ingress} and \texttt{hynoc\_local\_interface}),
\texttt{sclkfifolut} (single-clock variant, used in single-clock mode), and
\texttt{prra} together with its \texttt{prra\_lut} sub-module (the parallel
round-robin arbiter, used in \texttt{hynoc\_egress}).
These components are independently verified and reusable outside HyNoC.
Relying on shared library primitives rather than vendor-specific FIFO or
arbitration IP is what makes the design portable across FPGA families.

\subsection{RTL Organization}

The HyNoC RTL is written in Verilog and organized as follows:

\begin{itemize}
  \item \texttt{hynoc\_ingress}: ingress port (dual-clock FIFO via
        \texttt{dclkfifolut}/\texttt{sclkfifolut}, routing FSM, unicast and
        multicast protocol decoder);
  \item \texttt{hynoc\_egress}: egress port (data multiplexer, PRRA via
        \texttt{prra}/\texttt{prra\_lut});
  \item \texttt{hynoc\_router\_base}: parameterized router base (crossbar wiring,
        port instantiation);
  \item \texttt{hynoc\_router\_3p} / \texttt{hynoc\_router\_5p}: 3-port and
        5-port router instances with self-checking testbenches;
  \item \texttt{hynoc\_local\_interface}: local interface for node attachment
        (output FIFO via \texttt{dclkfifolut}/\texttt{sclkfifolut});
  \item \texttt{hynoc\_stream\_reader} / \texttt{hynoc\_stream\_writer}:
        simulation traffic generators used in testbenches to inject and
        check random packet streams.
\end{itemize}

The design is parameterized by: payload width (\texttt{PAYLOAD\_WIDTH}), FIFO
depth (\texttt{LOG2\_FIFO\_DEPTH}: 2 to 64 entries), number of ports
(\texttt{NB\_PORTS}: 3 to 9), index width (\texttt{INDEX\_WIDTH}), optional
multicast routing (\texttt{ENABLE\_MCAST\_ROUTING}), and single-clock mode
(\texttt{SINGLE\_CLOCK\_ROUTER}).

\subsection{Simulation and Verification}

Each module is delivered with a self-checking testbench. The 3-port router
testbench exercises both unicast and multicast scenarios over a 2-router, 4-node
topology. The 5-port router testbenches cover five distinct traffic scenarios,
including simultaneous multi-path transfers that verify the PRRA's starvation-free
behavior.

Simulation is supported under Icarus Verilog and ModelSim. Verilator is
integrated in the build infrastructure both for linting (\texttt{--lint-only})
and for C++ co-simulation: a multi-instance Verilator model is used in
Section~\ref{sec:evaluation} to characterize end-to-end latency and link
utilization on a $4\times4$ mesh.  Waveform inspection is
automated using Wavedisp~\cite{wavedisp}, a Python module that generates
simulator-specific signal configuration scripts from a single source description,
targeting GTKWave, ModelSim, and Riviera-PRO.

\subsection{FPGA Target}

Table~\ref{tab:synth-k7} reports post-place-and-route results on two Xilinx
7-series devices: a Kintex-7 (xc7k160t-fbg484, speed grade~$-2$) targeting
370\,MHz and a Zynq-7020 (xc7z020clg484, speed grade~$-1$) targeting 200\,MHz.
Both runs use 32-bit payload and 32-entry single-clock FIFOs
(\texttt{sclkfifolut}).  \textbf{WNS} (Worst Negative Slack) is the tightest
setup slack across all register-to-register timing paths; a positive value
means the target is met, and $f_\text{max} = 1\,/\,(T_\text{clk} - \text{WNS})$.

\begin{table}[h]
\centering
\footnotesize
\setlength{\tabcolsep}{2pt}
\caption{P\&R results on two Xilinx 7-series devices (32-bit,
         32-entry \texttt{sclkfifolut}; 24\,LUT-RAM/port for ingress FIFOs).
         WNS in ns; $f_\text{max}$ in MHz.}
\label{tab:synth-k7}
\begin{tabular}{@{}l rr r rr rr@{}}
\toprule
& & & &
  \multicolumn{2}{c}{\shortstack{Kintex-7\;$-2$\\(370\,MHz)}} &
  \multicolumn{2}{c}{\shortstack{Zynq-7020\;$-1$\\(200\,MHz)}} \\
\cmidrule(lr){5-6}\cmidrule(lr){7-8}
Router & LUT & LUT & FFs & WNS & $f_\text{max}$ & WNS & $f_\text{max}$ \\
       & (log.) & (RAM) & & (ns) & (MHz) & (ns) & (MHz) \\
\midrule
3-port &  253 &  72 &  327 & $+$0.079 & 381 & $+$0.034 & 201 \\
5-port &  596 & 120 &  590 & $+$0.127 & 389 & $+$0.101 & 204 \\
7-port & 1259 & 168 &  889 & $+$0.070 & 380 & $+$0.032 & 201 \\
9-port & 1806 & 216 & 1215 & $+$0.003 & 371 & $+$0.034 & 201 \\
\bottomrule
\end{tabular}
\end{table}

LUT-RAM usage is exactly 24 per port in every configuration, reflecting the
fixed-depth FIFO at each ingress; no block RAM is consumed.  LUT counts are
device-independent to within 1\% (synthesis tool variation).  Logic LUTs grow
superlinearly with port count, driven by the PRRA whose table size scales as
$2^{N-1}$ per egress port.  FFs grow nearly linearly (109 to 135 per port),
consistent with the per-port register cost of the ingress FSM and FIFO control.
All four port counts meet their respective frequency targets on both devices,
with the 5-port configuration consistently achieving the best timing margin.

The design has been validated on both Intel (Altera) and AMD (Xilinx) FPGA
families; reference board support is included for the DE0-Nano (Cyclone~IV)
and the ZedBoard (Zynq-7000).  Both \texttt{dclkfifolut} (dual-clock) and
\texttt{sclkfifolut} (single-clock) map exclusively to distributed LUT RAM,
avoiding dependency on vendor-specific FIFO IP and ensuring portability across
FPGA families.

The AXI-Stream-compatible port interface (see Section~\ref{sec:architecture})
simplifies integration with standard FPGA IP cores. On Zynq devices for instance,
the HyNoC local interface connects directly to the Processing System AXI-Stream
DMA engine with only a thin adapter to bridge the \texttt{fifo\_level} sideband.
This reduces integration effort and allows reuse of existing AXI-Stream
verification infrastructure.

\subsection{RTL Elaboration with Veriparse}

HyNoC makes use of advanced Verilog constructs — in particular, \texttt{initial}
blocks containing \texttt{for} loops and integer arithmetic — to pre-compute the
PRRA lookup tables at elaboration time. While this is perfectly legal Verilog, some
FPGA toolchains have limited support for such constructs in synthesizable RTL.

Veriparse~\cite{veriparse} is a companion source-to-source transformation toolkit
that addresses this issue. Its \texttt{veriflat} tool performs a sequence of
elaboration-time passes on the full module hierarchy: \emph{parameter and
localparam inlining} substitutes all parameter values at instantiation sites;
\emph{constant folding} and \emph{expression evaluation} reduce symbolic
expressions to integer literals; \emph{loop unrolling} expands \texttt{for},
\texttt{while}, and \texttt{repeat} constructs; \emph{generate-block elaboration}
resolves \texttt{generate}/\texttt{endgenerate} sections; \emph{variable folding}
evaluates \texttt{initial} blocks whose control flow is entirely determined by
inlined constants; and \emph{dead-code elimination} prunes branches that can never
be taken after constant propagation. The \emph{module flattener} then inlines the
full instantiation hierarchy into a single, parameter-free module. The resulting
RTL contains only plain assignments, registers, and combinational logic, with no
remaining parametric expressions — portable across any standard Verilog toolchain.

Figure~\ref{fig:veriflat-router} illustrates parameter inlining and constant
folding on the module boundary: the parameterized \texttt{hynoc\_router\_5p}
declaration, whose port widths are symbolic expressions of \texttt{PAYLOAD\_WIDTH}
and \texttt{LOG2\_FIFO\_DEPTH}, becomes a flat declaration with all widths
resolved to integer literals. The 5-port, 32-bit configuration yields
\texttt{[32:0]} data buses and \texttt{[5:0]} FIFO level signals in the output.

\begin{figure}[t]
\textbf{Before (parameterized):}
\begin{lstlisting}
module hynoc_router_5p
  #(parameter PAYLOAD_WIDTH   = 32,
    parameter LOG2_FIFO_DEPTH = 5, ...)
   (...
    input  wire [PAYLOAD_WIDTH:0]   port0_ingress_data,
    output wire [LOG2_FIFO_DEPTH:0] port0_ingress_fifo_level,
    ...);
\end{lstlisting}
\textbf{After \texttt{veriflat} (parameters inlined):}
\begin{lstlisting}
module hynoc_router_5p
   (...
    input  wire [32:0] port0_ingress_data,
    output wire [5:0]  port0_ingress_fifo_level,
    ...);
\end{lstlisting}
\caption{\texttt{veriflat} resolves parameterized port widths into constants.}
\label{fig:veriflat-router}
\end{figure}

Figure~\ref{fig:veriflat-prra} shows a complementary transformation class
supported by \texttt{veriflat}: variable folding combined with loop unrolling.
The \texttt{initial} block of \texttt{prra\_lut} — which computes the
round-robin priority table through nested \texttt{for} loops and modular
arithmetic over inlined constants — is fully evaluated at elaboration time;
the resulting \texttt{initial} block contains only explicit constant assignments.

\begin{figure}[t]
\textbf{Before (algorithmic \texttt{initial} block):}
\begin{lstlisting}
initial begin
  lut[0] = STATE_OFFSET[LOG2_WIDTH-1:0];
  for (j=1; j<lut_length; j=j+1) begin
    lut_index = j[WIDTH-1:0];
    value = -1;
    for (k=0; k<WIDTH; k=k+1) begin
      l = (k + STATE_OFFSET + 1) % WIDTH;
      if (lut_index[l] == 1'b1)
        if (value == -1) value = l;
    end
    lut[j] = value[LOG2_WIDTH-1:0];
  end
end
\end{lstlisting}
\textbf{After \texttt{veriflat} (pre-computed, \texttt{STATE\_OFFSET=0}):}
\begin{lstlisting}
initial begin
  lut[0]=2'd0; lut[1]=2'd0; lut[2]=2'd1; lut[3]=2'd1;
  lut[4]=2'd2; lut[5]=2'd2; lut[6]=2'd1; lut[7]=2'd1;
  lut[8]=2'd3; lut[9]=2'd3; lut[10]=2'd1; lut[11]=2'd1;
  lut[12]=2'd2; lut[13]=2'd2; lut[14]=2'd1; lut[15]=2'd1;
end
\end{lstlisting}
\caption{\texttt{veriflat} evaluates the \texttt{prra\_lut} \texttt{initial}
         block: nested loops and modular arithmetic are replaced by 16 explicit
         constant assignments (4-port arbiter, \texttt{STATE\_OFFSET=0}).}
\label{fig:veriflat-prra}
\end{figure}

The flattened output also benefits simulation: a single-module design with no
hierarchy reduces simulator overhead and can significantly speed up large testbench
runs. Veriparse is available at \url{https://github.com/cclienti/veriparse} under
the LGPLv3 license.

\subsection{License}

HyNoC has been in development since 2013, as reflected in the copyright headers
of every RTL source file. The source code was first published in 2019; the
repository was formally placed under the
\textbf{CERN-OHL-P~v2}~license~\cite{cernohl} in 2026. This permissive open
hardware license allows use, modification, and redistribution with no copyleft
obligation on the end product. The complete source, including RTL, testbenches,
documentation, and build infrastructure, is available at:
\url{https://github.com/cclienti/verilog-ip}.

%% file: sections/evaluation.tex
\section{Evaluation}
\label{sec:evaluation}

We present three Verilator~\cite{verilator}-based simulations on a $4\times4$
HyNoC mesh, each targeting a different evaluation goal:
\begin{enumerate}
  \item \textbf{Latency characterization.}  A lightweight GeMV
        (General Matrix-Vector multiplication, $\mathbf{y}=A\mathbf{x}$) with a
        $16\times4$ matrix produces clean per-hop measurements that establish
        the deterministic per-hop latency discussed in
        Section~\ref{sec:switching}.
  \item \textbf{LLaMA~3~8B FFN benchmark (single master).}  A full-scale
        GeMV at LLaMA~3~8B FFN up-projection dimensions
        ($14336\times4096$) with Q8\_0 weights and BF16 activations
        demonstrates the network under realistic AI inference workload.
  \item \textbf{Four-master throughput scaling.}  The same LLaMA~3~8B-scale
        benchmark is replicated across four corner masters, each owning a
        $2\times2$ quadrant, to quantify the speedup from distributed
        injection.
\end{enumerate}
All three benchmarks share the same 16-router $4\times4$ mesh and cover the
full communication cycle: data scatter from the master(s), computation at
worker nodes, and result gather.

\subsection{Experimental Setup}

\textbf{Network Topology.}

\begin{figure}[t]
  \centering
  \begin{tikzpicture}[
    rtr/.style={circle, draw, font=\scriptsize, inner sep=1pt, minimum size=22pt},
    mst/.style={circle, draw, fill=black!18, font=\scriptsize\bfseries, inner sep=1pt,
                minimum size=22pt},
    >=stealth,
    route/.style={->, line width=1.6pt, color=black!55}
  ]
  \def\S{1.6}
  % Horizontal links
  \foreach \row in {0,...,3}{
    \foreach \col in {0,...,2}{
      \draw[<->] ({\col*\S+0.16},{-\row*\S}) -- ({\col*\S+\S-0.16},{-\row*\S});
    }
  }
  % Vertical links
  \foreach \row in {0,...,2}{
    \foreach \col in {0,...,3}{
      \draw[<->] ({\col*\S},{-\row*\S-0.16}) -- ({\col*\S},{-\row*\S-\S+0.16});
    }
  }
  % All nodes
  \foreach \row in {0,...,3}{
    \foreach \col in {0,...,3}{
      \node[rtr] at ({\col*\S},{-\row*\S}) {\row,\col};
    }
  }
  % Master override
  \node[mst] at (0,0) {\textbf{M}};
  % Sample route (0,0)→(2,3): East×3, South×2
  \draw[route] (0,0) -- ({\S},0) -- ({2*\S},0) -- ({3*\S},0)
               -- ({3*\S},{-\S}) -- ({3*\S},{-2*\S});
  \end{tikzpicture}
  \caption{$4\times4$ mesh topology. \textbf{M} = master at $(0,0)$; remaining
           nodes are workers at $(r,c)$.  Bidirectional links are full-duplex
           (independent ingress and egress).  The gray arrow illustrates the XY
           route from the master to worker $(2,3)$: three East hops followed by two
           South hops, then a final local hop ($H=6$).}
  \label{fig:mesh4x4}
\end{figure}
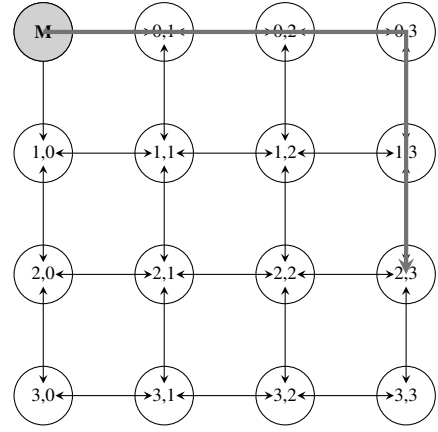

The network consists of 16~\texttt{hynoc\_router\_5p} instances arranged as
a $4\times4$ 2D mesh, as shown in Figure~\ref{fig:mesh4x4}.  Routers are
addressed by $(r,c)$ with $r\in\{0,1,2,3\}$ (row, top-to-bottom) and
$c\in\{0,1,2,3\}$ (column, left-to-right).  Port convention per router:
0\,=\,Local, 1\,=\,East, 2\,=\,South, 3\,=\,West, 4\,=\,North.  Each
router's local port (port~0) attaches a processing node; neighboring routers are
cross-linked on their directional ports (e.g., router $(r,c)$ port~1 connects to
router $(r,c{+}1)$ port~3).

The master node resides at $(0,0)$.  It sequentially distributes data to all 15
worker nodes and collects their results.

\subsection{Experiment 1: Latency Characterization}

\subsubsection{Computation and Communication Model}

The benchmark computes $\mathbf{y} = A\mathbf{x}$ with $A\in\mathbb{R}^{16\times4}$
and $\mathbf{x}=[1,2,3,4]^T$.  Worker $(r,c)$ is responsible for one scalar inner
product: $y_{4r+c} = \langle A_{4r+c},\, \mathbf{x} \rangle$.

\textbf{Forward packet} (master to worker $(r,c)$): one routing flit, one tag
flit, four flits for the matrix row $A_{4r+c}$, and four flits for
$\mathbf{x}$, totaling ten flits.  After the routing flit is consumed by the
network, the worker receives $P_\text{fwd}=9$ payload flits.

\textbf{Return packet} (worker to master): one routing flit, one tag flit, and one
result flit.  The master receives $P_\text{ret}=2$ payload flits.

Workers compute the inner product instantaneously, so the total simulation time
is dominated by communication latency and the sequential injection schedule at
the master.

\textbf{Route construction.}  Hop lists are built following dimension-order
(XY) order: East/West moves bring the packet to the target column, North/South
moves bring it to the target row, and a final local hop delivers it.  This
eliminates cyclic channel dependencies and guarantees deadlock
freedom~\cite{dally1987}.  All packets use standard unicast source routing
(proto \texttt{4'b0000}); the dedicated XY proto field (\texttt{4'b1000})
reserved in Table~\ref{tab:proto} is not exercised in these experiments.
A route from $(r_0,c_0)$ to $(r_1,c_1)$ requires
\begin{equation}
  H = |c_1-c_0| + |r_1-r_0| + 1
  \label{eq:hops-xy}
\end{equation}
hops, where the $+1$ accounts for the final local hop at the destination.
Throughout this section $H$ always counts all hops including that local hop.
For the master at $(0,0)$, this simplifies to $H=r+c+1$.  Routes are encoded as source-routing flits using the relative hop
encoding described in Section~\ref{sec:architecture}.

\textbf{Simulation Infrastructure.}

The Verilator C++ model instantiates 16~\texttt{Vhynoc\_router\_5p} objects
sharing one \texttt{VerilatedContext}.  Inter-router links are modeled as
one-cycle latches: egress signals are captured after each rising clock edge and
applied to the neighbor's ingress before the next rising edge.  This one-cycle
propagation delay is physically accurate for a registered synchronous
interconnect.

Each router drives a single clock (\texttt{router\_clk} and all
\texttt{portX\_ingress\_clk} toggled together).  In dual-clock mode
(\texttt{SINGLE\_CLOCK\_ROUTER=0}), the ingress dual-clock FIFOs internally
use gray-code CDC synchronization even when source and destination clocks are
the same signal; in single-clock mode (\texttt{SINGLE\_CLOCK\_ROUTER=1}), the
gray-code pipeline is bypassed.

\subsubsection{Latency Results}

Table~\ref{tab:latency} reports the \emph{return-path} latency for each hop
count $H$ in both configurations.  Return-path measurements are free of
source-queuing effects (each worker sends exactly one result as soon as
computation completes), yielding clean linear fits.

\begin{table}[h]
\centering
\footnotesize
\setlength{\tabcolsep}{4pt}
\caption{Measured return-path latency ($P=2$, cycles) vs.\ hop count.
         Dual = dual-clock FIFO (\texttt{SINGLE\_CLOCK\_ROUTER=0});
         Single = single-clock FIFO (\texttt{SINGLE\_CLOCK\_ROUTER=1});
         $\Delta$ = Dual$-$Single.}
\label{tab:latency}
\begin{tabular}{@{}rrrr@{}}
\toprule
$H$ & Dual & Single & $\Delta$ \\
\midrule
2 & 25 & 19 &  6 \\
3 & 37 & 28 &  9 \\
4 & 49 & 37 & 12 \\
5 & 61 & 46 & 15 \\
6 & 73 & 55 & 18 \\
7 & 85 & 64 & 21 \\
\bottomrule
\end{tabular}
\end{table}

The measured return-path latencies satisfy exact linear fits:
\begin{align}
  L_\text{dual}   &= 12H + P - 1 \label{eq:lfit-dual}   \\
  L_\text{single} &= 9H + P - 1  \label{eq:lfit-single}
\end{align}
matching the linear, deterministic form of~(\ref{eq:latency}) with a fixed
per-hop cost of $\alpha=12$ cycles in dual-clock mode and $\alpha=9$ in
single-clock mode, and no load-dependent term.  This per-hop cost is the
pipeline traversed at each router---ingress FIFO read, routing-FSM request
register, PRRA grant, and egress data register---together with the one-cycle
registered inter-router link of the co-simulation harness.  The
$3$-cycle-per-hop difference between the two modes ($\alpha=12$ versus $9$) is
exactly the depth of the dual-clock FIFO's gray-code
synchroniser~\cite{cummings2002}, which the single-clock FIFO does not
instantiate.  This difference is constant across all hop counts, so the latency
stays perfectly linear and predictable in both modes.

The total simulation time is \textbf{391~cycles} in dual-clock mode and
\textbf{349~cycles} in single-clock mode ($-$10.7\%).  All 16 computed
results are correct in both runs.

\subsubsection{Link Utilization}

Table~\ref{tab:links} summarizes link utilization over the 391-cycle dual-clock run.
Traffic is inherently asymmetric: the master sends all forward packets via its East
port first, concentrating load on the East-bound links of row~0.

\begin{table}[h]
\centering
\footnotesize
\setlength{\tabcolsep}{4pt}
\caption{Link utilization (\%) by directed link (dual-clock, 391~cycles).
         Each row names one router-to-router link.
         Links with $<$0.5\% utilization omitted; overall average: 3.3\%.}
\label{tab:links}
\begin{tabular}{@{}llr@{}}
\toprule
Direction & Link & Util\,(\%) \\
\midrule
\multirow{3}{*}{$\to$East, row\,0}
  & $(0,0)\!\to\!(0,1)$ & 30.7 \\
  & $(0,1)\!\to\!(0,2)$ & 20.5 \\
  & $(0,2)\!\to\!(0,3)$ & 10.2 \\
\cmidrule(l){1-3}
\multirow{3}{*}{$\to$South, col\,0}
  & $(0,0)\!\to\!(1,0)$ &  7.7 \\
  & $(1,0)\!\to\!(2,0)$ &  5.1 \\
  & $(2,0)\!\to\!(3,0)$ &  2.6 \\
\cmidrule(l){1-3}
$\to$North, col\,0
  & $(1,0)\!\to\!(0,0)$ &  9.2 \\
\cmidrule(l){1-3}
\multirow{3}{*}{$\to$West, row\,0}
  & $(0,1)\!\to\!(0,0)$ &  2.3 \\
  & $(0,2)\!\to\!(0,1)$ &  1.5 \\
  & $(0,3)\!\to\!(0,2)$ &  0.8 \\
\bottomrule
\end{tabular}
\end{table}

East-bound utilization halves at each column of row~0, consistent with XY
routing: each router delivers one packet to its local worker and forwards the
rest eastward.  The overall average of 3.3\% reflects the sequential master
bottleneck: with a single sender issuing one packet at a time, most links are
idle.

\subsection{Experiment 2: LLaMA~3~8B FFN Benchmark}

\subsubsection{Single-Master Configuration}
\label{sec:large-matmul}

To assess HyNoC under a realistic AI inference workload we scale the benchmark
to the LLaMA~3~8B~\cite{llama3} FFN up-projection layer.  In transformer
models, the Feed-Forward Network (FFN) consists of two linear projections with
a nonlinearity in between; the \emph{up-projection} is the first of these,
expanding the representation from the model dimension $D_\text{in}=4096$ to
the intermediate dimension $D_\text{out}=14336$.  The benchmark computes
$\mathbf{y}=W\mathbf{x}$ with $W\in\mathbb{R}^{14336\times4096}$ and
$\mathbf{x}\in\mathbb{R}^{4096}$.

Weights are stored in the Q8\_0 quantization format used by
llama.cpp~\cite{llama_cpp}: each weight is an 8-bit signed integer, and one
FP16 (16-bit floating-point) scale factor is shared per block of 32 weights to
recover approximate FP32 values at decode time.  Mapping to flits: the FP16
scale occupies the lower 16~bits of one 32-bit flit (upper 16~bits zero), and
the 32 int8 weights are packed four per flit, requiring 8~flits --- hence
9~flits per Q8\_0 block.  Activations use the BF16 format (Brain
Float~16: 1~sign bit, 8~exponent bits, 7~mantissa bits), packed two values per
32-bit flit.  Results are accumulated and returned in FP32 (IEEE~754 single
precision).

\textbf{Forward packet} (master to worker, for one output row): one routing
flit, one tag flit, 128~Q8\_0 blocks~$\times$~(1~scale flit + 8~weight flits),
and 2048~BF16 activation flits, totaling $P_\text{fwd}=3201$ payload flits.

\textbf{Return packet}: one routing flit, one tag flit, and one FP32 result
flit --- $P_\text{ret}=2$ payload flits, identical to the latency experiment.

\textbf{Computation model.}  Each of the 15~worker nodes is assigned
$14336/16 = 896$ output rows; the master at $(0,0)$ also computes 896~rows
locally.  Workers decode Q8\_0 weights to FP32 and accumulate the dot product
in FP32.  Local computation is modeled as instantaneous: injecting one forward
packet takes 3202~clock cycles (one per flit), which already exceeds the
dot-product budget of a real VLIW unit over 4096~elements, so arithmetic
is not the bottleneck.  All
14\,336~results are verified for exact FP32 equality against a software
reference.

\textbf{Throughput.}  The simulation completes in \textbf{43\,102\,246~cycles}
(dual-clock FIFO, \texttt{SINGLE\_CLOCK\_ROUTER=0})
with the master's single egress port sustaining near-100\% injection throughout.
Link utilization (Table~\ref{tab:links-llama}) shows the consequence of this
sustained injection: link $(0,0)\!\to\!(0,1)$ reaches 79.9\%,
carrying packets bound for all 12~workers that need at least one East hop.
Utilization falls in proportion to the number of remaining destinations on
each successive link.  South-bound links absorb $\tfrac{1}{4}$ of the total
traffic each.  Return-path links (West and North) are negligible at $<$0.1\%.
The overall average of \textbf{6.7\%} reflects the single-port injection
bottleneck: 47~of 48~directed links are underutilised while one port
saturates.  Repeating the run in single-clock mode
(\texttt{SINGLE\_CLOCK\_ROUTER=1}) yields \textbf{43\,102\,204~cycles}, only
42~cycles below the dual-clock result.  This confirms that the workload is
\emph{injection-bound} rather than latency-bound: runtime is dominated by the
one-flit-per-cycle injection of the 3201-flit forward packets, so the per-hop
latency saved by single-clock operation---the $9H$ versus $12H$ slope
of~(\ref{eq:lfit-single}) and~(\ref{eq:lfit-dual})---is negligible against the
tens of millions of injection cycles.  The four-master run is likewise
unchanged (8\,620\,466~cycles single-clock versus 8\,620\,484 dual-clock), so
the $5\times$ speedup is mode-independent.  Single-clock operation benefits
only latency-dominated traffic, such as the Experiment~1 latency test, where it
reduced the total from 391 to 349~cycles.

\begin{table}[h]
\centering
\footnotesize
\setlength{\tabcolsep}{4pt}
\caption{Link utilization (\%) by directed link, LLaMA~3~8B single-master run
         (43\,102\,246~cycles).  All four South columns are symmetric.
         Return-path links ($<$0.1\%) omitted; overall average: 6.7\%.}
\label{tab:links-llama}
\begin{tabular}{@{}llr@{}}
\toprule
Direction & Link & Util\,(\%) \\
\midrule
\multirow{3}{*}{$\to$East, row\,0}
  & $(0,0)\!\to\!(0,1)$ & 79.9 \\
  & $(0,1)\!\to\!(0,2)$ & 53.2 \\
  & $(0,2)\!\to\!(0,3)$ & 26.6 \\
\cmidrule(l){1-3}
\multirow{3}{*}{$\to$South, all cols}
  & $(0,k)\!\to\!(1,k)$, $k=0\text{--}3$ & 20.0 \\
  & $(1,k)\!\to\!(2,k)$, $k=0\text{--}3$ & 13.3 \\
  & $(2,k)\!\to\!(3,k)$, $k=0\text{--}3$ &  6.7 \\
\bottomrule
\end{tabular}
\end{table}

The load on each segment is strictly proportional to the number of destinations
reachable through it (East: 12, 8, 4 workers; South per column: 3, 2, 1
workers), with no hot-spots beyond the first East segment.  Nonetheless, the
bottleneck remains the single injection port, motivating the multi-master
extension.

\subsubsection{Four-Master Quadrant Configuration}
\label{sec:4master}

The single master at $(0,0)$ creates an injection bottleneck: all 15 forward
packets per round enter through one port, and all result packets return to
that same port.  We now distribute injection across the four corner nodes by
assigning each a $2\times2$ quadrant.

Table~\ref{tab:quadrants} lists the assignment.  Master~$m$ handles
$Q = D_\text{out}/4 = 3584$ output rows.  With $N_W=3$ workers per quadrant
and one row computed locally, each master executes
$\mathrm{ROUNDS}=Q/(N_W+1)=896$ \emph{rounds}---where one round consists of
sending one forward packet to each of the $N_W$ workers and computing one row
locally, then collecting $N_W$ result packets---the same count as the
single-master case, but all four masters proceed simultaneously.

\begin{table}[h]
\centering
\footnotesize
\caption{Quadrant assignment: masters, workers, and row ranges.}
\label{tab:quadrants}
\begin{tabular}{@{}clll@{}}
\toprule
Master & Position & Workers & Rows \\
\midrule
$M_0$ & $(0,0)$ & $(0,1),(1,0),(1,1)$ & $0$--$3583$ \\
$M_1$ & $(0,3)$ & $(0,2),(1,3),(1,2)$ & $3584$--$7167$ \\
$M_2$ & $(3,0)$ & $(3,1),(2,0),(2,1)$ & $7168$--$10751$ \\
$M_3$ & $(3,3)$ & $(3,2),(2,3),(2,2)$ & $10752$--$14335$ \\
\bottomrule
\end{tabular}
\end{table}

The structural property exploited here is \emph{link disjointness}: every XY
route from a corner master to its three workers stays within the $2\times2$
quadrant.  No packet crosses the horizontal midline (between rows~1 and~2) or
the vertical midline (between columns~1 and~2), so the four quadrants are
traffic-isolated by construction, without any runtime coordination.

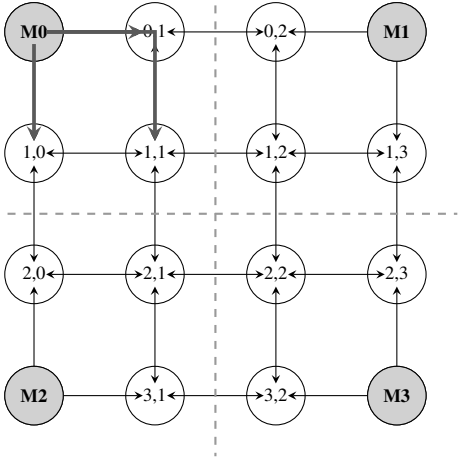
\begin{figure}[t]
  \centering
  \begin{tikzpicture}[
    rtr/.style={circle, draw, font=\scriptsize, inner sep=1pt, minimum size=22pt},
    mst/.style={circle, draw, fill=black!18, font=\scriptsize\bfseries, inner sep=1pt,
                minimum size=22pt},
    >=stealth,
    route/.style={->, line width=1.4pt, color=black!65}
  ]
  \def\S{1.6}
  % Horizontal links
  \foreach \row in {0,...,3}{
    \foreach \col in {0,...,2}{
      \draw[<->] ({\col*\S+0.16},{-\row*\S}) -- ({\col*\S+\S-0.16},{-\row*\S});
    }
  }
  % Vertical links
  \foreach \row in {0,...,2}{
    \foreach \col in {0,...,3}{
      \draw[<->] ({\col*\S},{-\row*\S-0.16}) -- ({\col*\S},{-\row*\S-\S+0.16});
    }
  }
  % Quadrant boundary (dashed)
  \draw[dashed, black!40, thick]
    ({1.5*\S},{0.35}) -- ({1.5*\S},{-3.5*\S});
  \draw[dashed, black!40, thick]
    ({-0.35},{-1.5*\S}) -- ({3.5*\S},{-1.5*\S});
  % All nodes
  \foreach \row in {0,...,3}{
    \foreach \col in {0,...,3}{
      \node[rtr] at ({\col*\S},{-\row*\S}) {\row,\col};
    }
  }
  % Corner masters
  \node[mst] at (0,0)              {\textbf{M0}};
  \node[mst] at ({3*\S},0)        {\textbf{M1}};
  \node[mst] at (0,{-3*\S})       {\textbf{M2}};
  \node[mst] at ({3*\S},{-3*\S})  {\textbf{M3}};
  % Routes from M0 to its three workers
  \draw[route] (0.16,0) -- ({\S-0.16},0);
  \draw[route] (0,-0.16) -- (0,{-\S+0.16});
  \draw[route] (0.16,0) -- ({\S},0) -- ({\S},{-\S+0.16});
  \end{tikzpicture}
  \caption{Four-master $4\times4$ mesh.  Corner nodes $M_0$--$M_3$ (shaded)
           each coordinate a $2\times2$ quadrant; dashed lines mark the mesh
           midpoints.  Arrows show $M_0$'s three routes to workers $(0,1)$,
           $(1,0)$, and $(1,1)$.  By symmetry, routes from all four masters
           remain within their respective quadrant.}
  \label{fig:mesh4x4-4m}
\end{figure}

Figure~\ref{fig:mesh4x4-4m} illustrates the topology and the three routes
from $M_0$: one East hop to $(0,1)$, one South hop to $(1,0)$, and
East+South to $(1,1)$.

\emph{Computation model.}  Each master computes one row locally per round
(row $mQ + t(N_W+1)$ in round~$t$, for $t = 0,\ldots,895$), for a total
of 896 local rows per master and 3584 across all four.  Local dot products
are modeled as instantaneous: a VLIW unit over $D_\text{in}=4096$ elements
completes well within the $3\times3202$ clock-cycle injection window per round,
so arithmetic is not on the critical path.  All 14\,336 results are verified
for exact FP32 equality against a software reference.

\subsubsection{Throughput Scaling Results}

The four-master simulation completes in \textbf{8\,620\,484~cycles}, a
$\mathbf{5.0\times}$ speedup over the 43\,102\,246~cycles of the
single-master run (Table~\ref{tab:throughput}).

\begin{table}[h]
\centering
\footnotesize
\setlength{\tabcolsep}{4pt}
\caption{Single- vs.\ four-master throughput
         ($D_\text{out}=14336$, $P_\text{fwd}=3201$ payload flits).}
\label{tab:throughput}
\begin{tabular}{@{}lrrr@{}}
\toprule
Configuration & Masters & Cycles & Speedup \\
\midrule
Single master & 1 & 43\,102\,246 & $1\times$ \\
Four masters  & 4 &  8\,620\,484 & $5\times$ \\
\bottomrule
\end{tabular}
\end{table}

The $5\times$ gain exceeds the $4\times$ expected from simple injection
parallelism.  The additional margin comes from backpressure relief: each
corner master serves only three workers, whereas the single master drives
fifteen.  With fewer concurrent result packets converging on one egress
port, injection stalls are shorter.

Table~\ref{tab:links4m} reports the utilization pattern.  The quadrant
isolation property is confirmed quantitatively: every link crossing either
mesh midpoint registers exactly 0\%.  This zero-overhead isolation is a
direct consequence of source routing: routes are computed statically and
encoded entirely in the packet header, so no router-internal mechanism is
needed to enforce quadrant boundaries at runtime.  Within each quadrant, the first East
or West hop from the corner master reaches 66.6\% because two of the three
forward routes share it (the direct worker and the diagonal worker); each
South or North second-hop segment carries one route at 33.3\%.  By symmetry
all four quadrants show identical figures.  The overall average of 11.1\%
contrasts with the 3.3\% of the single-master run, reflecting simultaneous
injection from four ports and correspondingly more uniform mesh loading.

\begin{table}[h]
\centering
\footnotesize
\setlength{\tabcolsep}{4pt}
\caption{Representative link utilization (\%), four-master run
         (8\,620\,484~cycles).  By symmetry all four quadrants are
         identical; cross-boundary links are exactly 0\%.}
\label{tab:links4m}
\begin{tabular}{@{}llr@{}}
\toprule
Direction & Link & Util\,(\%) \\
\midrule
$\to$East, row\,0  & $(0,0)\!\to\!(0,1)$ & 66.6 \\
$\to$West, row\,0  & $(0,3)\!\to\!(0,2)$ & 66.6 \\
$\to$East, row\,3  & $(3,0)\!\to\!(3,1)$ & 66.6 \\
$\to$West, row\,3  & $(3,3)\!\to\!(3,2)$ & 66.6 \\
$\to$South, col\,0 & $(0,0)\!\to\!(1,0)$ & 33.3 \\
$\to$South, col\,1 & $(0,1)\!\to\!(1,1)$ & 33.3 \\
$\to$North, col\,0 & $(3,0)\!\to\!(2,0)$ & 33.3 \\
$\to$North, col\,1 & $(3,1)\!\to\!(2,1)$ & 33.3 \\
Cross-boundary     & (any)               &  0.0 \\
\midrule
\multicolumn{2}{@{}l}{\textbf{Overall average}} & \textbf{11.1} \\
\bottomrule
\end{tabular}
\end{table}

All results in this section are obtained from Verilator co-simulation; on-device
measurement on physical FPGA hardware remains open and is identified as future
work in Section~\ref{sec:conclusion}.

%% file: sections/conclusion.tex
\section{Conclusion}
\label{sec:conclusion}

HyNoC is a hybrid circuit-switch/wormhole Network-on-Chip designed for distributed
VLIW computing on FPGA. By combining source routing---which establishes a dedicated
path through a distributed request/grant handshake---with wormhole flit streaming,
HyNoC achieves bounded, deterministic transfer latency without the area overhead of
virtual channels.

Congestion is managed statically by the compiler or system software, which selects
routes across the rich set of shortest paths available in mesh and
higher-dimensional topologies. The combinatorial analysis presented in
Section~\ref{sec:switching} shows that enriching the physical topology is more
area-efficient than adding virtual channels for the targeted workload class,
provided that a compile-time scheduler distributes traffic across available paths.

The parallel round-robin arbiter provides starvation-free, fixed-latency
arbitration at each egress port, and per-port independent clock domains allow the
NoC to interface cleanly with heterogeneous processing elements running at
different frequencies. The multiple local interface mechanism offers a practical
alternative to virtual channels when a node requires simultaneous non-blocking
communication with multiple peers.

Verilator co-simulation on a $4\times4$ mesh measures a deterministic per-hop
latency: return-path latency satisfies $L=12H+P-1$ (dual-clock) and
$L=9H+P-1$ (single-clock), the $3$-cycle-per-hop difference being exactly the
depth of the dual-clock FIFO's gray-code synchroniser, which single-clock
operation omits.  A
LLaMA~3~8B FFN up-projection benchmark (Q8\_0 weights, BF16 activations,
$14336\times4096$) with a single master completes in 43\,M cycles; a
four-master quadrant configuration, where each corner master owns a $2\times2$
traffic-isolated quadrant, reduces this to 8.6\,M cycles---a $5\times$ speedup
with zero cross-quadrant link traffic, confirming the deterministic
throughput scaling predicted by the link-disjoint topology design.

HyNoC is released as open-source hardware under the CERN-OHL-P~v2 license,
providing a complete RTL implementation with simulation testbenches and FPGA board
support.

Future work includes: formal deadlock-freedom analysis for arbitrary source route
sets; synthesis results and latency characterization on representative FPGA
devices (Section~\ref{sec:evaluation} provides Verilator-based characterization;
on-device measurement remains open); integration with a VLIW compiler back-end
for automatic route assignment and injection scheduling; and extension of the XY
routing protocol placeholder to a full hardware implementation.